\documentclass[11pt]{article}                                                                                
\usepackage{amsmath,amstext,amsbsy,amssymb}
\usepackage{bm}
\newcommand{\Tr}{{\rm Tr}\, }

\newcommand{\be}{\begin{equation}}
\newcommand{\ee}{\end{equation}}

\newcommand{\ssh}{{\scriptscriptstyle{1/2}}}

\newcommand{\ssK}{{\scriptscriptstyle{\rm K}}}

\newcommand{\ssR}{{\scriptscriptstyle{\rm R}}}

\newcommand{\ssCM}{{\scriptscriptstyle{\rm CM}}}
\newcommand{\ssCE}{{\scriptscriptstyle{\rm CE}}}
\long\def\symbolfootnote[#1]#2{\begingroup%
\def\thefootnote{\fnsymbol{footnote}}\footnote[#1]{#2}\endgroup}

\begin{document}

\begin{center}

{\Large \bf   

\vspace{18mm}

A Semiclassical Kinetic Theory of Dirac Particles and \\ Thomas Precession

}

\vspace{18mm}

\"{O}mer F. DAYI and Eda KILIN\c{C}ARSLAN

\vspace{5mm}

{\em {\it Physics Engineering Department, Faculty of Science and
Letters, Istanbul Technical University,\\
TR-34469, Maslak--Istanbul, Turkey}}\footnote{{\it E-mail addresses:} dayi@itu.edu.tr, edakilincarslann@gmail.com. }

\vspace{10mm}

\end{center}
  Kinetic theory of Dirac fermions is studied within the matrix valued differential forms method. It is based on the symplectic form derived by employing  the semiclassical wave packet build of the positive energy solutions of the Dirac equation. A satisfactory definition of the distribution matrix  elements imposes to work in the basis where the helicity is diagonal which is also  needed to attain the massless limit. We show that the kinematic  Thomas precession correction  can be studied straightforwardly within this approach. It contributes on an equal footing with the Berry gauge fields. In fact in equations of motion it eliminates the terms arising  from the Berry gauge fields.

\vspace{2cm}

%\newpage

\section{Introduction} 

Dirac  equation which describes massive spin-$1/2$  particles  possesses either positive or negative energy solutions described by 4-dimensional spinors. However, to furnish a well defined one particle interpretation, instead of employing  both type of solutions, a wave packet build  of only positive energy plane wave solutions should be preferred.
A nonrelativistic semiclassical dynamics  can be obtained employing this wave packet. 
Semiclassical limit may be useful to have a better understanding  of some quantum mechanical phenomena.  In the massless limit Dirac equation
leads to  two copies of Weyl particles which possess opposite chirality.
Recently, the chiral semiclassical kinetic theory has been formulated to embrace the anomalies due to the external electromagnetic fields 
 in $3+1$ dimensions \cite{soy,sy}. This remarkable result was extended to any even dimensional space-time by making use of differential forms in
\cite{ds} by introducing some classical variables corresponding to spin. Although in \cite{ds} non-Abelian anomalies have been incorporated into the particle currents, the solutions of phase space velocities in terms of phase space variables were missing.
In \cite{de} a complete description of the chiral semiclassical kinetic theory in any even dimension was established by introducing
 a  symplectic two-form which is a matrix labeled with ``spin indices", without introducing any  classical variable corresponding to  spin degrees of freedom.  In this formalism, although the classical phase space variables are the ordinary ones, the velocities arising from the equations of motion  are matrix valued.
 It has been shown in \cite{dy} that this matrix valued symplectic two form  can be derived within the semiclassical wave packet formalism 
\cite{sundaramniu, Culcer}. In fact the  ``spin indices" label the linearly independent  positive energy solutions.  

 In the formulation of the  Hamiltonian dynamics  starting with  a first order Lagrangian,  the related Hamiltonian  should be provided. In the development of the chiral kinetic theory  the Hamiltonian was taken as the positive relativistic energy of the free Weyl Hamiltonian. However, later it was shown that 
the adequate Hamiltonian should contain all the first  order terms in  Planck constant \cite{mtr} which can be attained by employing the method introduced in \cite{GBM}. Independently, the same Hamiltonian was conjectured in \cite{chenetal} to restore the Lorentz invariance of the semiclassical chiral theory. To obtain it  one first has to derive the Hamiltonian of the massive spin-$1/2$ fermion and then take the massless limit.

Massive fermions also appear in condensed matter systems which were studied in terms of  wave packets in \cite{chni,ccn}.
The semiclassical kinetic theory of  Dirac particles was also discussed in \cite{mwp}, where 
the Berry gauge fields have been given  in a different basis and some  classical  degrees of freedom have been assigned to spin. 
We would like  to establish the semiclassical kinetic theory of  Dirac particles within the formalism given in \cite{de}. There are some advantages of employing this method. First of all because of not attributing any classical variables to spin but taking them into account by considering  quantities matrix valued in ``spin space'', the calculations can be done explicitly. The differential forms method provides us the solutions of the equations of motion for the phase space velocities in terms of  phase space variables straightforwardly. Thus the particle currents can be readily derived. Moreover, we will show that within this formalism one can study the relativistic correction known as Thomas precession \cite{thomas1926}.

Thomas precession stems from the fact that a Lorentz boost can be written as two successive Lorentz boosts  accompanied by a rotation which is called Thomas rotation. This purely kinematic  phenomenon is essential to obtain the classical evolution of electron’s spin correctly, without referring to the Dirac equation. Thomas  precession should also contribute to the equations of motion of phase space variables. In fact, due to  Thomas precession the covariant formalisms of  Dirac particles yield equations of motion where anomalous velocity terms do not emerge \cite{hor,sdz}. However, as we will see   the equations of motion derived within the wave packet formalism possess anomalous velocity terms arising from the Berry curvature. This would have been expected because of the fact that our nonrelativistic formalism is not aware of  Thomas rotation. Correction due to Thomas rotation  should be installed in the formalism. We will show that our formalism suites well to take this correction into account: It contributes  to the 
one-form obtained by the semiclassical wave packet on an equal footing with the Berry gauge field. In fact,  it yields the cancellation of the anomalous velocity terms ignoring the  higher order terms  in  momentum. The connection of Berry gauge fields and Thomas precession was first observed by Mathur \cite{mathur}.

In Section \ref{packet} the one-form corresponding to the first order Lagrangian is obtained by the wave packet composed of the positive energy plane wave solutions of the Dirac equation. Then the related symplectic form is constructed and the solutions of the equations of motion for the velocities of the Dirac particle  are established in Section \ref{dynamics}.
Spin degrees of freedom are  taken into account by letting the velocities be matrix valued. So that, one should  consider a matrix valued distribution function. In contrary to to spin, helicity operator is a conserved quantity for the free Dirac particle.    Moreover,  we would like to split the particles as right-handed and left-handed appropriate to consider the massless limit and the chiral currents when there is an imbalance of chiral particles. Therefore we introduce a change of basis to the helicity basis as clarified in  Section \ref{helbas}.  Employing distribution matrix in the adequate basis we  then can write the particle number density and the related current by the velocities written in terms of the phase space variables in Section \ref{dynamics}. In Section \ref{0limit} we  obtained  the massless limit by  constructing the helicity eigenstates explicitly.
In Section \ref{Thomas}
a brief review  of  Thomas rotation is presented. Then,  we presented how it appears in the one-form obtained by the semiclassical wave packet. We will see that up to  higher order terms  in  momentum it contributes as the the Berry gauge field but with an opposite sign.
Our semiclassical formalism  should be supported by  an equation governing  spin dynamics. This will be attained employing   Gosselin-Ber\'{a}rd-Mohrbach (GBM)
method \cite{GBM} which is also needed to derive the semiclassical Hamiltonian.  In the last section the results obtained  and some possible applications  are discussed.
Moreover, we clarified  in which  frame  we obtained the semiclassical theory.

\section{Semiclassical wave packet}
\label{packet}

Dirac particle interacting with the external electromagnetic fields $\bm {{\cal E}},\ \bm B ,$ whose vector and scalar potentials are $\bm{a}(\bm{x})$ and $a_0(\bm{x}),$ is described by the Dirac Hamiltonian
$
H=H_0 + e a_0(x) ,$
where 
\begin{equation}
H_0(\bm{p}-e  \bm{a}(\bm{x}))=\beta m + \bm{\alpha} \cdot (\bm{p}- e \bm{a}(\bm{x})).
\label{hamiltonian}
\end{equation}
We set the speed of light $c=1$ and let $e<0$ for electron.
The representation of the $\alpha_i;\ i=1,2,3,$ and $\beta$ matrices are chosen as
$$
\alpha_i=
\begin{pmatrix}
 0 & \sigma_i \\
\sigma_i & 0
\end{pmatrix},\qquad
\beta=
\begin{pmatrix}
1 & 0\\
0 & -1
\end{pmatrix} ,
$$
where $\sigma_i$ are the Pauli spin matrices. 

A semiclassical formulation given in terms of the wave packet made of the positive energy solutions of the Dirac equation provides a well defined one particle interpretation. Let the position of the wave packet center in the coordinate space be $\bm{x}_c,$ and the corresponding momentum be 
$\bm{p}_c.$ The semiclassical wave packet is defined in terms of  the positive energy solutions, $u^\alpha (\bm{x}, \bm{p});$ $\alpha=1,2,$ as
$$
\psi_{\bm x} (\bm{p}_c,\bm{x}_c) = \sum_\alpha \xi_\alpha  u^\alpha  (\bm{p}_c,\bm{x}_c) e^{-i\bm{p}_c \cdot \bm{x} /\hbar}.
$$ 
For simplicity we deal with  constant $\xi_\alpha $ coefficients.
We would like to attain the one-form $\eta$ which is defined through $dS$ as
$$
dS \equiv \int [dx] \delta(\bm{x}_c - \bm{x}) \Psi^\dagger_{\bm{x}}\left( -i \hbar d - H_{0D} dt\right)\Psi_{\bm{x}}=\sum_{\alpha\beta}\xi^*_{\alpha} \eta^{\alpha\beta} \xi_{\beta}.
$$ 
$H_{0D}$ is the block diagonal Hamiltonian which should be derived from (\ref{hamiltonian}). Now, calculating  $dS$ one
attains the $\eta$ one-form as follows.
\begin{equation}
\eta^{\alpha\beta}= - \delta^{\alpha\beta}\bm{x}_c\cdot d\bm{p}_c - \bm a^{\alpha\beta}\cdot d{\bm x}_c - \bm A^{\alpha\beta}\cdot d{\bm p}_c - 
 H_{0D}^{\alpha\beta} dt .
\label{et1}
\end{equation}
Here we introduced the matrix valued Berry gauge fields
\begin{eqnarray}
\bm a^{\alpha\beta} &=& i \hbar u^{\dagger(\alpha)}(\bm p_c,\bm x_c)\frac{\partial }{\partial {\bm x_c}} u^{(\beta)}(\bm p_c,\bm x_c), \nonumber\\
\bm A^{\alpha\beta} &=& i \hbar u^{\dagger(\alpha)}(\bm p_c,\bm x_c)\frac{\partial }{\partial {\bm p_c}} u^{(\beta)}(\bm p_c,\bm x_c). \nonumber 
\end{eqnarray}

Although we deal with the $(3+1)$-dimensional space-time, the derivation of  $\eta$ does not depend on dimension.

\section{Semiclassical Hamiltonian dynamics }
\label{dynamics}

Instead of solving the Dirac equation in the presence of the electromagnetic vector potential $\bm{a}(\bm x),$
 we substitute 
$\bm{p} \rightarrow \bm{p}+e\bm{a}(\bm x),$  in (\ref{hamiltonian}) and  consider the free particle solutions with $E=\sqrt{p^2+m^2}.$ Then the positive energy solutions will not possess $\bm x$ dependence.
Therefore by renaming $(\bm{x}_c,\bm{p}_c)\rightarrow (\bm{x},\bm{p})$ and setting $\bm a^{\alpha\beta}=0,$ we obtain the  one-form
\begin{eqnarray}
\eta=p_i {dx}_i + e a_i {dx}_i - A_i {dp}_i - H dt.
\label{eta}
\end{eqnarray}
The repeated indices are summed over. We suppress the matrix indices $\alpha,\beta,$ and do not  write explicitly the unit matrix
 $\mathbb{I},$ when it is not necessary. 
 We deal with the Hamiltonian
$
H=H_D(\bm p) +ea_0({\bm x}),
$
where  
\begin{equation}
H_D(\bm p)= E - \hbar e \left( m \frac{\bm{\sigma} \cdot \bm{B}}{2 E^2} + \frac{(\bm{B} \cdot \bm p)(\bm{\sigma} \cdot \bm p)}{2E^2(E+m)}\right).
\label{hdh}
\end{equation}
This semiclassical Hamiltonian includes all contributions which are at the first order in $\hbar.$ It is attained by making use of the  GBM method \cite{GBM}. 

To obtain Hamiltonian dynamics we have to introduce  a symplectic two-form. In general the constituents of the one-form (matrix) 
$\eta$ can be non-Abelian, so that we adopt the definition of the symplectic two-form $\tilde{\omega}$ to be
\begin{equation}
\tilde{\omega} = d\eta - \frac{i}{\hbar} \eta \wedge \eta. \nonumber
\end{equation} 
We would like to emphasize that it is a matrix in spin indices.
Employing the one-form  (\ref{eta}), it yields
$$
\tilde{\omega} = {dp}_i \wedge{dx}_i + e{\cal E}_i {dx}_i \wedge dt + f_i{dp}_i \wedge dt - G + e F.
$$
Here, ${\cal E}_i =- \left(  \frac{\partial a_i}{\partial t} + \frac{\partial a_0}{\partial x_i}\right),$ is the electric field and 
\begin{equation}
f_i = -\frac{\partial H_D(\bm p)}{\partial p_i} - \frac{i}{\hbar} [A_i, H_D(\bm p)].
\label{fi}
\end{equation}
We always keep the terms up to the first order in $\hbar.$
The two-forms $F=\frac{1}{2} F_{ij}{dx}_i \wedge {dx}_j$  and $G=\frac{1}{2} {G_{ij}}{dp}_i \wedge {dp}_j$ 
are written in terms of
the electromagnetic field strength 
$$
F_{ij}=\left( \frac{\partial a_j}{\partial x_i}-\frac{\partial a_i}{\partial x_j}\right)=\epsilon_{ijk}B_k,
$$
and the Berry curvature  
\begin{eqnarray}
G_{ij} &=&\left( \frac{\partial A_j}{\partial p_i}- \frac{\partial A_i}{\partial p_j}-i[A_i,A_j]\right)= {\epsilon}_{ijk}G_k
\label{eq:G}.
\end{eqnarray}

In order to derive the Berry gauge fields let us present the positive energy solutions for  $H_0(\bm p) =\bm{\alpha} \cdot \bm{p} + \beta m.$ They can be obtained as
$$
u^\alpha (\bm{p}) = U(\bm{p}) {u_0}^\alpha,
$$
where 
${u_0}^{1T}=(1\ 0\ 0\	0), \ \ {u_0}^{2T}=(0\ 1\ 0\	0),$  are the rest frame solutions and  $U(\bm{p})$ is the Foldy-Wouthuysen transformation:
$$
U(\bm{p}) =\frac{\beta H_0 (\bm p)+ E}{\sqrt{2E (E+m)}}.
$$ 
Now, the Berry gauge fields  can be written as
$$
A_i = i\hbar  I_+  U(\bm{p}) \frac{\partial U^\dagger(\bm p)}{\partial p_i}  I_+ .
$$
$I_+$ projects onto the positive energy subspace. Hence,  we acquire
\begin{equation}
\bm{A}=-\hbar \frac{\bm{\sigma} \times \bm{p}}{2E(E+m)}.
\label{eq:A}
\end{equation}
By making use of it in (\ref{eq:G}) one gets
\begin{equation}
\bm G= -\frac{\hbar m}{2E^3}\left( \bm{\sigma}+\frac{\bm{p}(\bm{\sigma}\cdot\bm{p})}{m(m+E)}\right)
\label{berrycurvature}.
\end{equation}
Observe that  the covariant derivative of $\bm G$ vanishes:
\begin{equation}
D_i G_i = \frac{\partial G_i}{\partial p_i} - \frac{i}{\hbar} [A_i,G_i]=0.
\label{covdef}
\end{equation}

To derive the equations of motion let us introduce the vector field
\begin{equation}
\label{vf}
\tilde v= \frac{\partial}{\partial t}+\dot{\tilde x}_i\frac{\partial}{\partial x_i}+\dot{\tilde p}_i\frac{\partial}{\partial p_i}.
\end{equation}
$(\dot{\tilde x}_i, \dot{\tilde p}_i)$ are the matrix-valued time evolutions of the phase space variables $(x_i, p_i).$
By demanding that the interior product of  the vector field  with the symplectic form satisfy
$$
i_{\tilde{v}} \tilde{\omega} = 0, 
%\label{i_v}
$$
the equations of motion are found to be
\begin{eqnarray*}
\dot{\tilde x}_i &= &f_i \ + \ \dot{\tilde p}_j \ G_{ji},\\
\dot{\tilde p}_i &= &e{\cal E}_i \ - \ e \ \dot{\tilde x}_j \ F_{ji}.
\end{eqnarray*}

Time evolution of the volume form
$$\tilde{\Omega}=\frac{1}{3!} \tilde{\omega} \wedge \tilde{\omega} \wedge \tilde{\omega} \wedge dt ,$$
will yield the Liouville equation. 
The volume form can be written in terms of the canonical volume element of the phase space $dV$ as
$$\tilde{\Omega}= \tilde{\omega}_\ssh \ dV \wedge dt.$$
Here $\tilde{\omega}_\ssh$ is the Pfaffian of the following $(6\times 6)$ matrix,
$$\begin{pmatrix}
	F_{ij} & -\delta_{ij}\\
	\delta_{ij} & G_{ij}
\end{pmatrix} $$
Obviously, $\tilde{\omega}_\ssh$ is still a matrix in the spin indices. We will present its explicit form.

The Liouville equation can be provided by calculating the Lie derivative of the volume form associated with the vector field (\ref{vf}).
It can be calculated in  two different ways. First, the Lie derivative of the volume form can be written in terms of the Pfaffian as
\begin{eqnarray}
{\cal{L}}_{\tilde v} \tilde{\Omega} &=& (i_{\tilde v}  d + d i_{\tilde v} ) (\tilde{\omega}_\ssh dV \wedge dt) \nonumber\\
&=& \left( \frac{\partial \tilde{\omega}_\ssh}{\partial t} + \frac{\partial ( \dot{\tilde x}_i\tilde{\omega}_\ssh)}{\partial x_i} + \frac{\partial (\tilde{\omega}_\ssh \dot{\tilde p}_i)}{\partial p_i}\right) dV \wedge dt.
\label{lievolume}
\end{eqnarray}
The second way is
\begin{eqnarray}
{\cal{L}}_{\tilde v} \tilde{\Omega} &=& (i_{\tilde v}  d + d i_{\tilde v} )(\frac{1}{3!} \tilde{\omega}^3 \wedge dt)\nonumber\\
&=& \frac{1}{3!} d {\tilde{\omega}}^3 .
\label{liouville}
\end{eqnarray}
Now, calculate (\ref{liouville}) explicitly and compare with  (\ref{lievolume}). This provides the Pfaffian as well as the velocities of the phase space variables in terms of the phase space variables themselves:
\begin{eqnarray}
\tilde{\omega}_\ssh &=&1 - e \frac{\hbar m}{2E^3}\left( \bm{\sigma} \cdot \bm{B}+\frac{(\bm{p} \cdot \bm{B}) (\bm{\sigma}\cdot\bm{p})}{m(m+E)}\right),
 \label{eqomegamassive} \\
 \dot{\tilde{\bm x}}\tilde{\omega}_\ssh &=&-\bm f -  \frac{\hbar e m}{2E^3}\left(\bm{{\cal E}} \times \bm \sigma 
+\frac{ (\bm{{\cal E}} \times \bm p )(\bm \sigma \cdot \bm p)}{m(m+E)}\right)- e \bm B (\bm f \cdot \bm G ),
\label{xdotmassive} \\
\tilde{\omega}_\ssh \dot{\tilde{\bm p}} &=&
e\bm{{\cal E}}- e \bm{f} \times \bm{B} -  \frac{\hbar e^2m}{2E^3}\left( \bm \sigma 
+\frac{\bm p (\bm{\sigma}\cdot\bm{p})}{m(m+E)}\right) (\bm{{\cal E}} \cdot \bm{B}).
\label{pdotmassive}
\end{eqnarray}
On the other hand the term which survives in the Lie derivative is
\begin{equation}
{\cal{L}_\nu} {\Omega} =  \hbar \frac{e^2m (\bm{\sigma} \cdot \bm{p}) (\bm{{\cal E}} \cdot \bm{B} )}{E^4 (E+m)}  dV \wedge dt .
\label{liuson}
\end{equation}
This  can be observed  by explicit calculations.

\section{Helicity basis and the continuity equation}
\label{helbas}

We study spin-$1/2$ massive particles where the velocities are matrices in spin indices. So that we should consider a matrix valued distribution function. For a Dirac particle spin is given by \mbox{
$
\bm \Sigma = \begin{pmatrix}
 \bm \sigma & 0 \\
0 & \bm \sigma
\end{pmatrix} .
$}
Hence, for the semiclassical wave packet composed of the positive energy solutions spin matrices become
${u^\alpha}^\dagger \ \bm{\Sigma} \ u^\beta = {\bm{\sigma}}^{\alpha \beta}.$
They do not commute with the free Dirac Hamiltonian, so that they are not conserved in time. However the helicity operator
$$ {\lambda}^{\alpha \beta}= {u^\alpha}^\dagger \ (\frac{\bm{\Sigma}\cdot \bm{p}}{p}) \ u^\beta = \frac{\bm{\sigma}\cdot \bm{p}}{p}$$
commutes with the free Dirac Hamiltonian.
It is appropriate to split up the particles as right-handed and left-handed which can be performed in the basis where the helicity  is diagonal. In order to establish the diagonal basis we use the spherical coordinates where
$\lambda$ is diagonalized by the unitary matrix
$$
R=\begin{pmatrix}
\cos(\frac{\theta}{2}) & -\sin(\frac{\theta}{2}) e^{-i \varphi}\\
\sin(\frac{\theta}{2}) e^{i \varphi} & \cos(\frac{\theta}{2})
\end{pmatrix} .
$$
Thus the helicity basis is defined by 
$$\phi= R  u.$$
Now, we can define  the distribution matrix   in the helicity basis as
$$f_\phi=\begin{pmatrix}
f_R & 0\\
0 & f_L
\end{pmatrix}.$$
Inverse transformation will lead to the distribution matrix in the initial basis:
$$
f_u=R f_\phi R^{\dagger} = \begin{pmatrix}
 f_R\cos^2\frac{\theta}{2} + f_L\sin^2\frac{\theta}{2}  & \frac{1}{2} \sin\theta  e^{-i\phi} (f_R - f_L) \\
\frac{1}{2}\sin\theta e^{i\phi} (f_R - f_L) & f_R \cos^2\frac{\theta}{2}  + f_L\sin^2\frac{\theta}{2} 
\end{pmatrix}.
$$
Since, for massive spin-$1/2$ particles the number of the right-handed and left-handed particles are equal we set $f_R=f_L\equiv f$. Therefore,  in the initial basis the distribution matrix  becomes 
$$
f_u=  f \mathbb{I} .
$$
Let the distribution function, $f,$ satisfy the collisionless Boltzmann equation:
\begin{equation}
\frac {df}{dt}=\frac{\partial f}{\partial t}+\frac{\partial f}{\partial x_i}\dot{\tilde{x}}_i+\frac{\partial f}{\partial p_i}\dot{\tilde{p}}_i=0.
\label{boltzmann}
\end{equation}

In order to write the continuity equation, one should identify the particle number density $n(x,p,t),$ and the particle current density $j(x,p,t)$. 
However, we should first give an appropriate definition of the classical limit. This is  done by taking the trace over spin indices and defining  the classical velocities as follows
$$
\sqrt{W}\equiv \Tr[\tilde{\omega}_\ssh],\ \sqrt{W} \dot{\bm x}\equiv \Tr[ \dot{\tilde{\bm x}}\tilde{\omega}_\ssh], \ \sqrt{W} \dot{\bm p}\equiv \Tr[\tilde{\omega}_\ssh \dot{\tilde{\bm p}}].
$$
We can write the probability density function as $\rho(x,p,t)= \sqrt{W} f.$ Hence, the particle number density and the particle current density are given by
$$
n(x,t) = \int \frac{d^3p}{(2\pi)^3} \Tr[\tilde{\omega}_\ssh] f , \ \ \ 
\bm j(x,t)= \int \frac{d^3p}{(2\pi)^3} \Tr[ \dot{\tilde{\bm x}}\tilde{\omega}_\ssh ]f.
$$
Employing (\ref{lievolume}) and (\ref{liuson}) and setting
 $\int \frac{d^3p}{(2\pi)^3} \ \frac{\partial (\tilde{\omega}_\ssh \dot{\tilde{p}}_i)}{\partial p_i} = 0,$ because we suppose that there is no contribution from the the momentum space boundary, we find
\begin{eqnarray}
\frac{\partial n}{\partial t} + \bm{\nabla} \cdot \bm{j} = \int \frac{d^3p}{(2\pi)^3} {\rm Tr}[\hbar \frac{m (\bm{\sigma} \cdot \bm{p})(e\bm{{\cal E}} \cdot \bm{B})}{E^4 (E+m)}  ] f = 0 . \nonumber
\end{eqnarray} 
Thus we established the continuity equation  for Dirac particles.

One can also obtain the particle current 
$$
\bm j = \int \frac{d^3p}{(2\pi)^3} \Tr [ \dot{\tilde{\bm x}}\tilde{\omega}_\ssh ]f = \int \frac{d^3p}{(2\pi)^3}  \frac{\bm p }{E}f.$$

\section{Massless Fermions}\label{0limit}

 The massless Dirac equation can be written as two copies of the Weyl equation one for right-handed and one for left-handed fermions. Therefore, the helicity basis is suitable to discuss the massless case.

First of all in the limit of vanishing mass, the  Berry curvature in the initial basis (\ref{eq:G}) yields
\begin{equation}
\bm{G}_0=-\hbar  \bm{b}  \frac{\bm{\sigma}\cdot\bm{p}}{p}.
\label{bczer}
\end{equation}
We introduced  $\bm b=\bm p /2 p^3$ which is the monopole field situated at the origin of  momentum space: 
\mbox{
$
\bm{\nabla} \cdot \bm b = 2 \pi {\delta}^3 (p).
$ }
As it has  already been  announced the helicity basis is appropriate to discuss the massless case. In fact,
when we change the basis from $u$ to $\phi$, the Berry curvature (\ref{bczer}) becomes
$$
\bm{G}_{0\phi} = R^\dagger \ \bm{G}_0 \ R =  - \hbar \bm{b} \begin{pmatrix}
1 & 0\\
0 & -1
\end{pmatrix}. 
$$
Now, the Berry curvature is singular and instead of (\ref{covdef}) it satisfies
$$
\frac{\partial G_{0\phi}^i}{\partial p_i}=-2\pi\hbar \sigma_z {\delta}^3 (p).
$$

One can deal with the right-handed and left-handed fermions independently. Let us consider the right-handed massless fermions
in terms of the projection
\begin{equation}
\label{rpro}
P_R = \begin{pmatrix}
1 & 0\\
0 & 0 
\end{pmatrix}.
\end{equation}
Projection of the Pfaffian, 
$P_R \ \tilde{\omega} P_R$ yields the scalar value
$$
\sqrt{\omega} = 1-e  \hbar \bm{b} \cdot \bm{B}.
$$
Projecting  (\ref{xdotmassive}) and (\ref{pdotmassive}) one acquires the solutions for the Weyl fermion as
\begin{eqnarray*}
\sqrt{\omega} \dot{\bm x} &=&-{\bm f}^\ssR -\hbar  e\bm{{\cal E}} \times  \bm b + e \hbar \bm B  (\bm{f}^\ssR \cdot \bm{b}),
%\label{xdotmassless},
\\
\sqrt{\omega} \dot{\bm p} &= &e\bm{{\cal E}} - e  {\bm f}^\ssR \bm B + e^2  \hbar \bm b  (\bm{{\cal E}} \cdot \bm{B}).
%\label{pdotmassless} 
\end{eqnarray*}
${\bm f}^\ssR$ denotes the massless limit of ${\bm f},$ (\ref{fi}), projected by (\ref{rpro}).

Let $f$ be  the distribution function of the right-handed fermions which satisfies the collisionless Boltzmann equation (\ref{boltzmann}). The particle current density $\bm j$ can be written as
$$
\bm j = \int \frac{d^3p}{(2\pi)^3} \sqrt{\omega} \dot{\bm x} f = \int \frac{d^3p}{(2\pi)^3} \left(-{\bm f}^\ssR -\hbar e  \bm{{\cal E}} \times \bm b +  \hbar e \bm B(\bm{f}^\ssR \cdot \bm{b})\right) f. 
$$
The last term where the current is parallel to the magnetic field is known as the chiral magnetic effect term.
The continuity equation becomes anomalous:
$$
\frac{\partial n(x,t)}{\partial t}+\bm{\nabla} \cdot \bm{j}=\frac{e^2}{4\pi^2}f(x,p=0,t)\bm{{\cal E}}\cdot\bm{B}.
$$
The Berry monopole situated  at $|\bm{p}|=0$ is responsible for the non-conservation of the chiral particle current.

\section{Thomas Precession}\label{Thomas}

We would like to clarify how to take into account  Thomas precession \cite{thomas1926} within the wave packet formalism as presented here. To this end let us first briefly recall  Thomas precession following \cite{jackson1975}. 
The source of this phenomenon lies in the fact that if one would like to  write a Lorentz boost as two  successive   Lorentz boosts, she  should also rotate the coordinates with an angle depending on the velocities. This rotation yields an angular velocity known as  Thomas precession. 

Suppose that a  particle is moving with the velocity $\bm v$ with respect to laboratory frame at time $t.$ Hence the particle's co-moving frame denoted by 
the inertial spacetime coordinates $x^{\prime}$, is connected to the spacetime coordinates of the laboratory frame $x,$ by the Lorentz boost ${\lambda_{boost}}(\bm v)$ at time $t:$
\begin{equation}
x^{\prime}= {\lambda_{boost}}(\bm v)  x.
\label{l1}
\end{equation} 
Let the particle accelerates, so that it moves with the velocity $\bm v+d \bm v$ with respect to the laboratory frame at time $t+d t$. Then at time $t+d t$ the co-moving coordinate frame coordinates denoted $x^{\prime \prime},$ will be connected to the laboratory frame $x,$ by the Lorentz transformation 
\begin{equation}
x^{\prime \prime}= {\lambda_{boost}}(\bm v+d \bm v)  x.
\label{l2}
\end{equation}
Now,
let us write the connection between the two co-moving frame coordinates $x^{\prime}$ and $x^{\prime \prime}$ as 
$$x^{\prime \prime}= \lambda_T x^{\prime}.
$$ 
Inspecting (\ref{l1}) and (\ref{l2}) the transformation $\lambda_T$ can be written as
\begin{eqnarray}
\lambda_T = {\lambda_{boost}}(\bm v+d \bm v) \ {\lambda_{boost}}(-  \bm v).
\label{lambdat}
\end{eqnarray}

The Lorentz boost ${\lambda_{boost}}(\bm v+d \bm v)$ can be separated into two successive Lorentz boosts accompanied by the rotation $R(d \bm{\theta})$
\begin{equation}
\lambda_{boost}(\bm v+d \bm v) = R(d \bm{\theta})  \lambda_{boost}(d \bm v) \lambda_{boost}(\bm v) .
\label{l3}
\end{equation}
The infinitesimal angle of  rotation is
\begin{equation}
d \bm \theta =\frac{\gamma^2}{\gamma +1} \bm v \times d\bm v .
\label{theta}
\end{equation}
$\gamma$ is the Lorentz factor.
Therefore by plugging (\ref{l3}) into (\ref{lambdat}) one obtains
\begin{eqnarray}
\lambda_T = R(d \bm{\theta})  {\lambda_{boost}}(d \bm v)
\label{lambdat2}.
\end{eqnarray}

When one deals with  nonrelativistic equations of motion it is desired that in contrary to $x^{\prime \prime},$ 
 successive co-moving frames of the  particle would be  connected  only by  boosts  without any rotation. Therefore, $x^{\prime \prime \prime}$ coordinates of the frame moving with the velocity  $\bm v +d \bm v,$ at time $t+d t,$ will be obtained from the system moving with the velocity  $\bm v ,$ at time $t,$ only with the boost $\lambda_{boost}(d \bm v)$ without any rotation:
\begin{equation}
x^{\prime \prime \prime} = {\lambda_{boost}}(d \bm v)  x^{\prime}. \nonumber
\end{equation}
Then by making use of (\ref{l1}) and (\ref{l3}) the coordinates of the co-moving frame, $x^{\prime \prime \prime}$, are written in terms of the laboratory frame coordinates as
\begin{equation}
x^{\prime \prime \prime} = R(-d \bm{\theta}) \ {\lambda_{boost}}( \bm v + d \bm v ) \ x.
\label{x3}
\end{equation}

How this phenomena manifests itself in the semiclassical kinetic theory of Dirac particles? In the wave packet formalism one deals with the group velocity
\begin{equation}
\bm v \equiv\frac{\partial E }{\partial \bm p}=\frac{\bm p}{E}.
\label{veg}
\end{equation}
Since $\gamma=E/m,$  by plugging (\ref{veg}) into (\ref{theta}) one gets
$$
d \bm \theta=\frac{\bm p\times d \bm p}{m(E+m)}.
$$
Let the laboratory frame and co-moving reference frames coincide at the time $t=0$ when the particle is at rest. Hence the solution of the Dirac equation in laboratory frame at  $t=0,$ is $u(0)$. Now, in the nonrelativistic kinetic theory formulation one deals with
\begin{eqnarray*}
{du(\bm p)}_{NR} &=& u^{\prime \prime \prime} (\bm p + d \bm p)- u^{\prime} (\bm p) \\
& =& R(-d \bm{\theta}) \ {\lambda_{boost}}( \bm v + d \bm v ) u(0)- {\lambda_{boost}}( \bm v) u(0) \nonumber\\
&=& R(-d \bm{\theta}) u(\bm p + d \bm p) - u(\bm p) .
\end{eqnarray*}
Here we ignore the terms at   order of $p^2, $  so that we write $d\bm v = d\bm p /m.$
Therefore,  when one considers  Thomas precession, the Berry gauge field related term in the $\eta$ one-form (\ref{et1}),  of the nonrelativistic  formalism will be
\begin{equation}
i \hbar u^\dagger (\bm p) {du(\bm p)}_{NR} = i \hbar u^\dagger (\bm p)\left[R (-d \bm \theta) u (\bm p +d\bm p) -u (\bm p)\right]
\label{adtho}
\end{equation}
The infinitesimal rotation of spinors is given by
$$
R (d \bm \theta) = 1-\frac{i}{4}  \sigma_{ij}  {d \omega}^{ij} \equiv 1 + {\cal D} (d \bm \theta)
$$ 
We implicitly work with the metric $g={\rm diag} (1,-1,-1,-1),$ so that  we have
\begin{equation}
\sigma_{ij}  = \epsilon_{ijk} \begin{pmatrix}
	\sigma_k & 0 \\
	0 & \sigma_k 
\end{pmatrix} , \ \ 
{d \omega}^{ij} =\epsilon^{ijm}  {d \theta}_m = - \epsilon^{ijm}  {d \theta}^m .
\end{equation}
Therefore, the rotation is expressed by
$$
{\cal D} (d \bm \theta)=\begin{pmatrix}
\bm{\sigma} \cdot d \bm{\theta} & 0 \\
0 &  \bm{\sigma} \cdot d \bm{\theta} \nonumber \\
\end{pmatrix}. \nonumber
$$ 
Then, keeping only the first order terms in $d\bm p ,$ (\ref{adtho}) yields
$$
{i \hbar u^\dagger (\bm p)  d u (\bm p)}_{NR}=i \hbar u^\dagger (\bm p)\frac{\partial u (\bm p)}{\partial \bm p}\cdot d \bm p 
+i \hbar u^\dagger (\bm p)   {\cal D} (-d \bm \theta) u (\bm p).
$$
 The first term is the Berry gauge field calculated in (\ref{eq:A}). The Thomas correction term can be shown to be
\begin{equation}
i \hbar u^\dagger (\bm p)   {\cal D} (-d \bm \theta) u (\bm p)= \frac{( \hbar \bm \sigma \times \bm p )}{4m^2}\cdot d \bm p.
\label{AT}
\end{equation}
Ignoring the terms at the order of $p^2$ the Berry gauge fields  (\ref{eq:A}) yield the same contribution as in (\ref{AT}), up to a minus sign. 
Therefore, when the Thomas correction is considered the velocities can be read from
(\ref{eqomegamassive}), (\ref{xdotmassive}) and (\ref{pdotmassive}) by setting the  Berry gauge fields to zero. We conclude that the anomalous velocity terms disappeared when  Thomas rotation is taken into account. This is in accord with the results obtained in relativistic formulations of  Dirac particles \cite{hor,sdz}.

\newcommand{\ssa}{{\scriptscriptstyle{+}}}

\section{Time Evolution of Spin}
\label{bmt}
In the semiclassical approach adopted in this work, there is no classical degrees of freedom corresponding to spin of the particle.
However, for the sake of completeness we would like to supply  the time evolution equation of spin matrices.
In Section \ref{helbas} we found that for the wave packet composed of the positive energy solutions spin is  given by the Pauli spin matrices, $\bm{\sigma}.$

To furnish the semiclassical equation governing time evolution of spin we may employ the GBM method \cite{GBM}.
In the GBM method one deals with the semiclassical phase space coordinates which are noncommuting:
\begin{equation}
\bm{r}=  \bm x - \hbar \frac{\bm{\sigma} \times \bm{p}}{2E(E+m)} .
\label{ncoco}
\end{equation}
Now, we can define the time evolution of spin by
\begin{equation}
\frac{d \bm \sigma}{dt}= \frac{1}{ i  \hbar} [ \bm\sigma , H_D(\bm p)+ea_0(\bm r)] = \frac{1}{ i  \hbar}  \left( [\bm \sigma , H_D(\bm p)]
-e[\bm \sigma,r_i] \frac{\partial a_0 (\bm x)}{\partial x_i} \right).
\label{sigmadot}
\end{equation}
We deal with time independent electromagnetic potentials, so that $(\partial a_0/ \partial \bm x )=\bm{{\cal E}}.$ Moreover,  
in the Hamiltonian (\ref{hdh}) we ignore the higher order terms in $p^2:$
\begin{equation}
H_D(\bm p)=E - \frac{e  \hbar \bm{\sigma} \cdot \bm{B}}{2E} .
\label{BMTH}
\end{equation}
By plugging (\ref{ncoco}) and (\ref{BMTH}), into (\ref{sigmadot}) and setting $\gamma=E/m,$ one establishes the time evolution of  spin:
\begin{eqnarray}
\frac{d \bm{\sigma}}{dt} = \frac{e}{m} \bm{\sigma} \times\left[\frac{1}{\gamma}\bm B +\frac{1}{\gamma +1}\bm E\times \bm v\right]. 
\label{BMT}
\end{eqnarray}
Observe that (\ref{BMT}) is the Bargmann-Michel-Telegdi equation \cite{BMT} as composed in  \cite{jackson1975}. 

\section{Discussion}

Semiclassical kinetic theory of massive spin-$1/2$ particles is studied within the method where the symplectic two-form  is a matrix in the so called  spin indices which label the positive energy solutions of the Dirac equation.  These solutions are the building blocks of  the wave packet which leads to the semiclassical approximation. Differential forms are employed to define the semiclassical Hamiltonian dynamics of  Dirac particles. Solutions of the equations of motion for  phase space velocities in terms of  phase space variables are derived. To define the continuity equation of the particle number density and particle number current one has to define the distribution matrix adequately. This is possible in the basis where the helicity operator is diagonal. Therefore, a change of basis  is performed. This is also needed  to establish the massless limit. In the massless limit the expected  anomalous continuity equation as well as the particle current possessing  the chiral magnetic effect are obtained. 

Thomas precession correction needed in the nonrelativistic formulation of dynamics is studied within the wave packet formalism. We showed that up to higher order terms in momentum it sweeps out the contributions arising from the Berry gauge fields. This coincides with the results obtained in relativistic formulation of  Dirac particles. This is the main result obtained in this work. The presented method of introducing Thomas precession correction   is valid in general. It can be applied to other semiclassical approaches of  Dirac like systems where the underlying Hamiltonian of the theory is given by Dirac like Hamiltonian as in some condensed matter systems. 

We built the semiclassical kinetic theory of Dirac particles starting with the wave packet. This formalism  is valid when the potentials do not vary rapidly across the wave packet. Now, the particle is  an extended object whose diameter is equal to the Compton wavelength \cite{ccn}. It is a self-rotating wave packet, so that its center of mass and  center of charge or the ``center of energy,"  do not coincide \cite{ghw}.  However, the  circular current which yields this interpretation is derived from the equations of motion for  the center of mass \cite{chni}. Thus, in the absence of  the electromagnetic vector field $\bm a =0,$  kinetic part of the one-form (\ref{eta}) can be written in the symmetric  form:
\begin{equation}
\eta_\ssK=\frac{1}{2}\left({\bm p}_{\ssCM} \cdot  d{\bm   x}_{\ssCM} -{\bm x}_{\ssCM} \cdot d {\bm p}_{\ssCM}\right)- \bm A \cdot d {\bm p}_{\ssCM} .
\label{kof}
\end{equation}
As we have already observed in Section \ref{helbas}, spin of the wave packet is given by $\bm S =\frac{\hbar}{2}\bm \sigma$. Thus the Berry gauge field can be expressed as 
$$
\bm A=-\frac{1}{E+m}\bm S \times \bm v\approx -\frac{1}{2m}\bm S \times \bm v,
$$
by keeping the terms linear in the velocity $\bm v .$
On the other hand  for a spinning object the center of mass and the center of energy are related as \cite{ghw} 
\begin{eqnarray}
{\bm x}_{\ssCM} &\approx &{\bm x}_{\ssCE}+ \frac{1}{m} \bm S \times \bm v = {\bm x}_{\ssCE} -2 \bm A ,\\
{\bm p}_{\ssCM} & \approx & {\bm p}_{\ssCE}.
\end{eqnarray}
Therefore, in terms of the center of energy variables $({\bm x}_{\ssCE} , {\bm p}_{\ssCE}),$  the kinetic one-form (\ref{kof}),  can be rewritten in the canonical form:
\begin{equation}
\eta_\ssK\approx\frac{1}{2}\left({\bm p}_{\ssCE} \cdot  d{\bm   x}_{\ssCE} -{\bm x}_{\ssCE} \cdot d {\bm p}_{\ssCE}\right).
\end{equation}
The Thomas precession affects the center of mass  and leaves intact the center of energy \cite{ghw}. Phase space variables of the GBM method refers to the center of energy. In fact because of this one can directly take the vanishing mass limit \cite{bmpla}.
Moreover, as we showed in Section \ref{bmt},  the Hamiltonian density   (\ref{hdh}), which depends only on the momenta ${\bm p}_{\ssCE}  \approx {\bm p}_{\ssCM},$
engenders the correct time evolution of spin without adding a Thomas precession term.     

The semiclassical kinetic theory formulation of Dirac particles and obtaining the massless limit by constructing the suitable basis which we  developed here, can be generalized to systems where some other interaction terms are present  in the underlying Hamiltonian. 

\newpage
\newcommand{\PRL}{Phys. Rev. Lett. }
\newcommand{\PRB}{Phys. Rev. B }
\newcommand{\PRD}{Phys. Rev. D }

%\end{document}

\end{document}